\documentclass[10pt, aps, pra, twocolumn, superscriptaddress, longbibliography, floatfix]{revtex4-1}

\usepackage[T1]{fontenc}
\usepackage[utf8]{inputenc}
\usepackage[english]{babel}
\usepackage{setspace}
\usepackage{comment}
\usepackage[dvipsnames]{xcolor}
\usepackage{enumitem}
\usepackage[colorlinks, bookmarks=false, citecolor=NavyBlue, linkcolor=red, urlcolor=blue]{hyperref}

\usepackage{amsmath, amssymb, amsfonts, mathrsfs}
\usepackage{mathtools}
\usepackage{physics, braket}
\usepackage{verbatim}
\usepackage{bbold, bm}
\usepackage[bbgreekl]{mathbbol}
\usepackage{nicefrac}
\usepackage{tipa}
\usepackage{wasysym}

\usepackage{tabularray}
\usepackage{multirow}
\usepackage{graphicx}
\usepackage[abs]{overpic}
\usepackage{booktabs}
\graphicspath{{figures/}}

\usepackage{tikz}
\usetikzlibrary{arrows}
\usetikzlibrary{math}
\usetikzlibrary{calc}
\usetikzlibrary{shapes.geometric}

\newcommand{\padA}{\affiliation{Dipartimento di Fisica e Astronomia “G. Galilei”, Universit{\`a} degli Studi di Padova, via Marzolo 8, I-35131 Padova, Italy.}}
\newcommand{\padB}{\affiliation{Padua Quantum Technologies Research Center, Universit{\`a} degli Studi di Padova, Italy.}}
\newcommand{\padC}{\affiliation{INFN, Sezione di Padova, Italy.}}
\newcommand{\iovA}{\affiliation{Medical Physics Department, Veneto Institute of Oncology IOV-IRCCS, Padova, Italy.}}
\newcommand{\saclay}{\affiliation{Université Paris-Saclay, CNRS, LPTMS, 91405, Orsay, France.}}
\newcommand{\planqc}{\affiliation{PlanQC GmbH, Lichtenbergstr. 8, 85748 Garching, Germany.}}

\usepackage{orcidlink}
\newcommand{\orcidsamuele}{\orcidlink{0000-0002-4706-1551}}
\newcommand{\orcidmarco}{\orcidlink{0009-0000-5194-3445}}
\newcommand{\orcidilaria}{\orcidlink{0000-0002-3806-2034}}
\newcommand{\orcidsimone}{\orcidlink{0000-0002-8882-2169}}
\newcommand{\orcidalessandro}{\orcidlink{0000-0002-1933-5658}}
\newcommand{\orcidmarta}{\orcidlink{0000-0001-8647-911X}}
\newcommand{\orcidgianluca}{\orcidlink{0009-0003-4073-7314}}

\definecolor{smoothred}{HTML}{C5232F}
\definecolor{mygreen}{rgb}{0,0.5,0}
\definecolor{myblue}{rgb}{0,0,0.75}
\definecolor{mymagenta}{cmyk}{0,1,0,0.12}

\newcommand{\costfunc}{F}
\newcommand{\nvars}{N_{\scriptscriptstyle V}}
\newcommand{\genericdomain}{\mathcal{D}}
\newcommand{\nqubits}{N_{\scriptscriptstyle Q}}
\newcommand{\normalizationcte}{K}

\newcommand{\refvalue}{a}
\newcommand{\discrstep}{\delta}
\newcommand{\genericbit}{b}
\newcommand{\historystep}{t}

\newcommand{\sgvar}{s}
\newcommand{\ham}{\mathcal{H}}
\newcommand{\sgbias}{h}
\newcommand{\sgcoupling}{J}

\newcommand{\funcevaluation}{n_{\scriptscriptstyle \mathrm{evaluation}}}
\newcommand{\maxfuncevaluation}{n^{\scriptscriptstyle \mathrm{max}}_{\scriptscriptstyle \mathrm{evaluation}}}
\newcommand{\nsamples}{n_{\scriptscriptstyle \mathrm{samples}}}

\newcommand{\dosedistribution}{\mathrm{D}}
\newcommand{\nbeamlets}{N_{\scriptscriptstyle B}}
\newcommand{\influencematrix}{\mathcal{A}}
\newcommand{\influencemel}{\influencematrix}
\newcommand{\nvoxel}{\nu}

\newcommand{\bonddimension}{m}

\begin{document}

\title{Adaptive Qubit Encoding for Continuous-Variable Quantum Optimization}

\author{Samuele Cavinato\orcidsamuele} \iovA
\author{Marco Tesoro\orcidmarco} \padA \padB \padC \planqc
\author{Gianluca Morettini\orcidgianluca} \saclay
\author{Marta Paiusco\orcidmarta} \iovA
\author{Alessandro Scaggion\orcidalessandro} \iovA
\author{Simone Montangero\orcidsimone} \padA \padB \padC
\author{Ilaria Siloi\orcidilaria} \padA \padB \padC


\begin{abstract}

\noindent Optimizing continuous-variable functions on qubit-based quantum hardware requires a finite representation of the search space and a solver-compatible representation of the objective function. 
Fixed-grid discretizations provide a direct encoding, but restrict the search to a predefined range and resolution; at the same time, constructing a compact Hamiltonian for a general continuous-variable objective can be impractical. 
Here, we introduce a framework for continuous-variable optimization tailored to qubit-based quantum architectures. 
The method combines an adaptive discretization with an iterative local mapping of the objective function onto an effective spin-glass Hamiltonian. 
We assess the approach on unconstrained and constrained instances of the $n$-dimensional Gramacy-Lee benchmark, and on a proof-of-concept inverse-planning problem in intensity-modulated radiotherapy (IMRT) for cancer treatment.
The results show that this framework can produce competitive solutions, supporting its use as a flexible mapping strategy for continuous-variable optimization in qubit-based quantum and quantum-inspired pipelines.

\end{abstract}
\maketitle


Continuous-variable optimization is at the core of a broad range of scientific and technological tasks, from parameter estimation and inverse problems to optimal control and engineering design~\cite{nocedal2006_continuous-optimization, polyak2021_continuous-optimization}. 
At a general level, these problems require the minimization of a real-valued objective function over continuous decision variables, where the structure of the objective function is defined by the target application~\cite{Abbas_2024}. 

A large class of quantum and quantum-inspired optimization methods encodes the objective function into a finite-dimensional cost Hamiltonian, whose low-energy configurations represent candidate solutions.
This formulation underlies quantum annealing and several variational or quantum-inspired optimization schemes.
The most common models are Ising Hamiltonians and their binary-variable equivalents, such as Quadratic Unconstrained Binary Optimization (QUBO) formulations; higher-order polynomial binary objectives are often written as Polynomial Unconstrained Binary Optimization (PUBO) models and, when required, reduced to quadratic form through quadratization or auxiliary variables~\cite{kochenberger2004_qubo, kochenberger2014_qubo-survey, lucas2014_np-probs-as-ising, punnen2022_main-qubo-ref, glover2022_quantum-bridge-analytics}. 
For continuous-variable optimization, this requires two additional modeling choices: how to encode the continuous search space into finite qubit registers, and how to express the objective function, exactly or approximately, as a Hamiltonian acting on those registers.

A standard approach is to discretize each continuous variable on a fixed grid and encode the corresponding grid index in binary form~\cite{arai2023_continuous-variable_qa}.
While compact in terms of qubits~---~$\nqubits$ qubits represent $2^{\nqubits}$ admissible values~---~optimization range and resolution need to be fixed before the search.
Increasing the resolution requires additional qubits, while an inadequate range may prevent the optimizer from accessing relevant regions of the continuous domain.
Alternative encodings~---~e.g. Gray-code, unary, one-hot, domain-wall, and nonuniform discretizations~---~improve locality, constraint handling, or resolution allocation~\cite{dominguez2023_cop-encoding-4-qc, dimatteo2021_gray-encoding, chang2022_gray-encoding}, but all of these alternatives fundamentally rely on a predefined finite representation of the continuous variables.

Native continuous-variable quantum approaches provide an alternative route, in which real-valued degrees of freedom are represented directly by bosonic modes rather than by qubit registers~\cite{verdon2019_cv-qaoa, matwiejew2023_cv-qaoa, arai2023_continuous-variable_qa, enomoto2023_exp_cv-qaoa}. 
While this avoids imposing a fixed discrete grid on the variables, here we focus on qubit-based architectures, where the relevant task is to construct a compact finite-dimensional representation of the continuous search space.

Once the continuous variables have been discretized, any objective function over bit-strings can, in principle, be represented as a PUBO. 
In practice, however, the resulting polynomial may be dense, high-order, or require interpolation, lookup-table constructions, auxiliary variables, or quadratization before it can be used in standard Ising/QUBO workflows~\cite{mandal2020_pubo1, domino2022_pubo2, salehi2022_pubo3, norimoto2023_pubo4}. 
For non-polynomial, non-differentiable, or numerically defined objectives, the construction of a compact global polynomial model can therefore become a significant bottleneck.

To address these issues, we introduce a framework based on two components.
The first is an adaptive discretization, in which the reference value of each continuous variable is encoded by a small qubit register with a variable-dependent resolution parameter.
The second is the iterative construction of local surrogate model, in which the objective function is sampled on neighboring configurations and locally matched to an effective spin-glass Hamiltonian.
The resulting model is quadratic in the spin variables and can therefore be processed by quantum or quantum-inspired solvers. We optimize the surrogate Hamiltonians with a tree tensor network (TTN) ansatz through a DMRG-like variational ground-state search~\cite{silvi2019_anthology}.

We assess the method on three classes of problems.
First, we consider the unconstrained $n$-dimensional Gramacy--Lee function, which provides a non-convex, oscillatory benchmark for continuous optimization~\cite{gramacy2012_glfunction}.
Second, we study a constrained variant of the same function, used to test the incorporation of penalty terms within the proposed framework. 
Third, we apply the method to a proof-of-concept fixed-gantry intensity-modulated radiotherapy (IMRT) inverse-planning problem, where the objective combines polynomial dose-matching terms with non-polynomial dose-constraint penalties~\cite{taylor2004_imrt, bortfeld2006_imrt}.
The TTN-based implementation is compared with a Best Neighbor baseline, Simulated Annealing applied to the same discretized surrogate Hamiltonian, and Particle Swarm optimization acting directly on the original continuous variables.

The remainder of the paper is organized as follows.
In Section~\ref{sec:M&M}, we introduce the dynamically adapted discretization, the local approximation model, the tensor-network optimization strategy, and the benchmark solvers. 
In Section~\ref{sec:results}, we present the numerical results for the benchmark problems and for the radiotherapy application. Finally, Section~\ref{sec:conclusions} summarizes the main findings, discusses the current limitations of the present study, and outlines possible future developments.

\section{Materials and methods}
\label{sec:M&M}

A general scalar continuous-variable optimization problem can be written as~\cite{nocedal2006_continuous-optimization, polyak2021_continuous-optimization}
\begin{equation}
    \label{eq:non_convex_problem}
    \vec{x}_{opt}
    = 
    \mathop{\arg\min}_{\vec{x} \in {\genericdomain}^{\nvars}} \,
    \costfunc(\vec{x}) \, ,
\end{equation}
where $\costfunc(\vec{x})$ is a scalar objective function that may be non-convex, non-differentiable, and non-polynomial.
The decision variables of the problem are collected in the vector $\vec{x} = \left(x_1, \ldots, x_{\nvars}\right)$, where $\nvars \in \mathbb{N}$ is the number of variables, and $\vec{x}_{opt}$ represents the optimal solution.
For the purposes of this work, we assume that each $x_j$ is real, i.e., $\genericdomain \equiv \mathbb{R}$.

Mapping this class of problems onto quantum hardware involves a two-step process:
\begin{enumerate}
    \item Each variable $x_j \in \mathbb{R}$ must be associated with a set of $\nqubits \in \mathbb{N}$ qubits that will encode its values during the optimization;
    \item The objective function $\costfunc(\vec{x})$ must be rewritten in a form compatible with the types of functions that can be represented on quantum hardware. 
\end{enumerate}
In the following Sections, we introduce an efficient framework that addresses both of these key requirements for continuous-variable quantum optimization.

\subsection{Dynamically adapted discretization}\label{subsec:dinamically-adapted}

The number of qubits $\nqubits$ assigned to a real-valued variable $x_j \in \mathbb{R}$ determines the number of discrete values that can be represented by qubit-based encodings~\cite{dominguez2023_cop-encoding-4-qc}. Under a binary encoding, $x_j$ can take one of $2^{\nqubits}$ possible values.
  
This standard binary discretization introduces two limitations in the representation of continuous variables: the encoding is restricted to a bounded domain, and the discretization cannot adapt to emerging information about the optimization landscape.
\begin{enumerate}
    \item \textbf{Bounded domain.}
    The variable $x_j= (1/\normalizationcte) \sum_{n=1}^{\nqubits} 2^{n - 1} \genericbit_n $, with $\genericbit_n \in \{0,1\}$   is restricted to the interval $\left[0, \left(2^{\nqubits} - 1\right) \mathbin{/} \normalizationcte\right]$, where $\normalizationcte \in \mathbb{R}$ is a normalization constant.
    An affine transformation can shift and rescale this domain so that $x_j \in \left[x_j^{\scriptscriptstyle \mathrm{min}}, \, x_j^{\scriptscriptstyle \mathrm{max}}\right]$;
    \item \textbf{Fixed discretization.}
    Within $\left[x_j^{\scriptscriptstyle \mathrm{min}}, \, x_j^{\scriptscriptstyle \mathrm{max}}\right]$, the admissible values of $x_j$  are determined by the initial discretization.
    In this standard approach, the discrete grid points remain unchanged throughout the optimization process.
    Consequently, there is in general no guarantee that the chosen discrete search space contains the global minimizer $\vec{x}_{opt}$ of the original continuous problem. 
\end{enumerate}
Beyond the standard binary representation, several alternative encodings have been proposed~\cite{dominguez2023_cop-encoding-4-qc}.
To name a few, Gray encoding~\cite{dimatteo2021_gray-encoding, chang2022_gray-encoding} modifies the ordering of binary strings so that consecutive values differ by only one bit, reducing sensitivity to bit-flip errors; however, it retains the uniform discretization and bounded-domain structure.
Nonuniform discretization schemes~\cite{dominguez2023_cop-encoding-4-qc} allocate basis states according to problem-specific priors or variable sensitivities, concentrating resolution in regions of greatest relevance; nonetheless, they require prior knowledge of the landscape and complicate operator construction.
Unary and one-hot encodings~\cite{dominguez2023_cop-encoding-4-qc} circumvent binary scaling altogether by representing values through the position of a single-excited qubit. This approach enhances interpretability and mitigates discretization artifacts, but increases the number of qubits $\nqubits$ required.
Finally, continuous-variable quantum representations~\cite{verdon2019_cv-qaoa, matwiejew2023_cv-qaoa, arai2023_continuous-variable_qa} encode real-valued variables directly, thereby eliminating discretization and fixed bounds in principle; yet they introduce practical challenges such as noise sensitivity, and the need for hardware beyond standard qubit architectures.

Each of these approaches relaxes some of the intrinsic constraints of binary encoding, yet none fully reconciles the trade-offs between representational flexibility, resource efficiency, and hardware feasibility.
To address these limitations, we propose an adaptive discretization framework.

Let us consider a decision variable $x_j$ with a reference value $\refvalue_j \in \mathbb{R}$.
We associate  $x_j$ with an $\nqubits$-qubit bit-string $\vec{\genericbit}^{\, (j)} = \left(\genericbit_{\nqubits}^{(j)} \ldots \genericbit_1^{(j)}\right)$.
For each $j \in \left[1, \ldots, \nvars\right]$, we introduce a \emph{discretization step} $\discrstep_j \in \mathbb{R}_{+}$, which determines the spacing between consecutive discrete possible values of $x_j$.
Specifically, the remaining $2^{\nqubits} - 1$ values are defined as integer multiples of $\discrstep_j$ taken around $\refvalue_j$.

\begin{figure}[t]
    \centering
    \begin{overpic}[width=1.0 \columnwidth, unit=1mm]{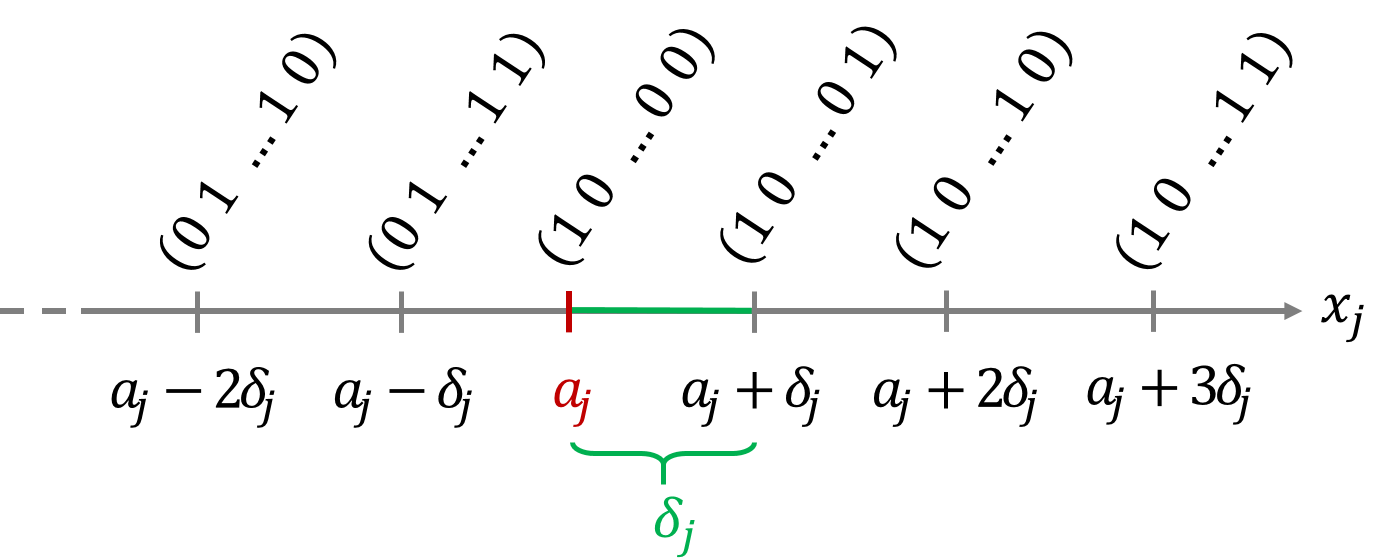}
    \end{overpic}
    \caption{
        \emph{Representation of the dynamically-adapted discretization.}
        The Figure illustrates the discretization procedure introduced in this work.
        Possible discretized values of a continuous decision variable $x_j \in \mathbb{R}$, represented on the real axis, are encoded using $\nqubits$-bit strings around a reference value $x_j = \refvalue_j$ (indicated by the red tick on the axis).
        Each variable $x_j$ is associated with its own dynamically adapted discretization parameter $\discrstep_j > 0$, which controls the range of values that can be encoded around the corresponding reference value $\refvalue_j$.
        Distinct bit-strings represent unique discretized values of $x_j$, obtained by adding or subtracting integer multiples of the discretization step $\discrstep_j$. 
    }
    \label{fig:discretization_scheme}
\end{figure}

For an $\nqubits$-bit binary sequence $\vec{\genericbit}^{\, (j)}$ (with qubits counted from right to left), we define
\begin{equation}
    \label{eq:flexible_discretization}
    x_j
    =
    \refvalue_j
    +
    \left(\sum_{n = 1}^{\nqubits} 2^{n - 1} \genericbit^{(j)}_n - 2^{\nqubits - 1}\right) \delta_j .
\end{equation}
This choice ensures that the sequence $\left(1 \, 0 \, 0 \, \ldots \, 0\right)$ reproduces the reference value $x_j = \refvalue_j$.
Note that the resulting discretization is not symmetric around $x_j = \refvalue_j$; the minimum and maximum deviations from $\refvalue_j$ are generally unequal. An example of this discretization scheme is illustrated in Figure~\ref{fig:discretization_scheme}. 

This strategy addresses the limitation on the \textbf{Bounded domain}.
To overcome the limitation on the \textbf{Fixed discretization}, we introduce a mechanism that  dynamically adapts the discretization step during the optimization process by leveraging information from the optimization history.
Specifically, the discretization step $\discrstep_j$ can be increased or decreased to explore alternative sets of values around a given configuration.
Since $\discrstep_j$ is labeled by the variable index $j \in \{1, 2, \ldots, \nvars\}$, each variable will evolve according to its own optimization history.
This refinement allows the available qubits to be used more effectively around a target configuration,
thereby improving the local precision of the representation. 

\begin{figure}[t]
    \centering
    \begin{overpic}[width=1.0 \columnwidth, unit=1mm]{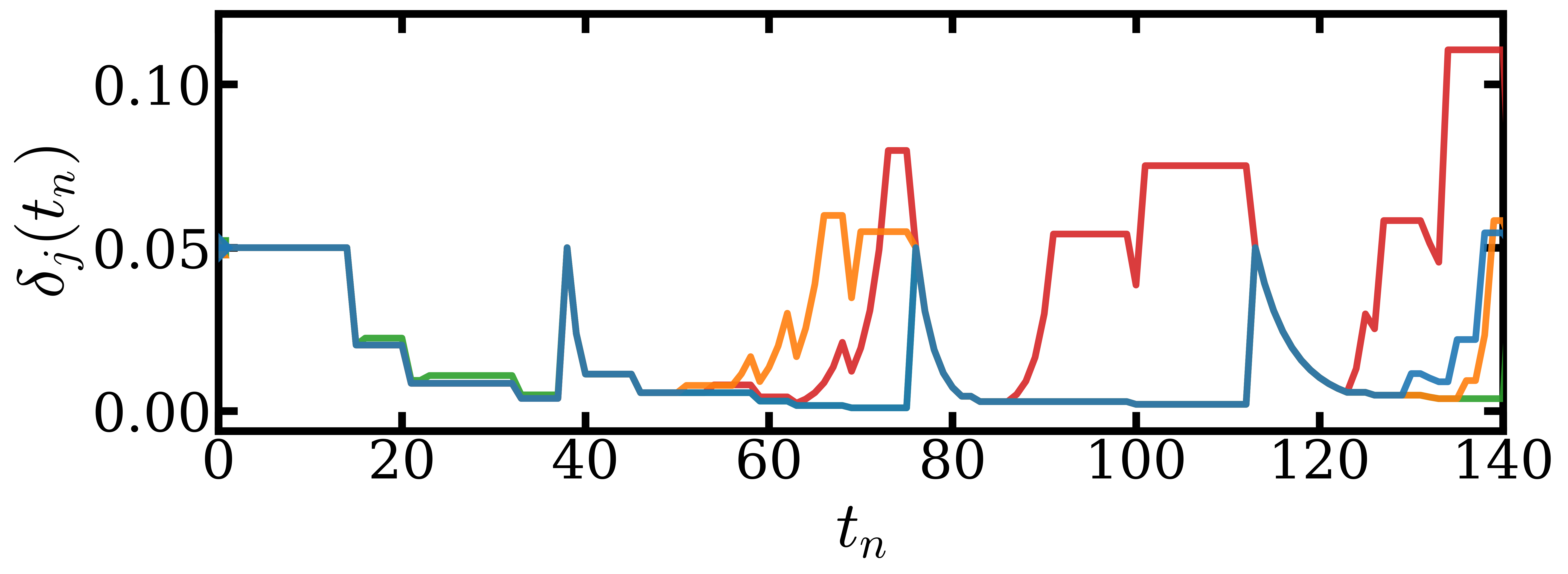}
    \end{overpic}
    \caption{
        \emph{Example of the variable-dependent evolution of the discretization step.}
        The red, green, blue and orange curves show the evolution of the step sizes $\discrstep_j\left(\historystep\right)$ associated with four representative variables $\left\{x_j\right\}_{j=1}^3$ over $140$ iterations of the optimization, all initialized at $\discrstep_i(0)=0.05$.
        As the algorithm progresses, the four step sizes evolve differently because each variable is updated according to its own local optimization history.
        The gradual reductions correspond to a refinement of the search region around the current reference value, while the abrupt increases mark update or refresh events that re-expand the region being explored.
        The Figure highlights the two key features of the proposed discretization scheme: its dynamic character and its variable-dependent adaptation.
    }
    \label{fig:sub-region_tn}
\end{figure}

The evolution of the discretization step $\discrstep_j > 0$ follows the schedule below:
\begin{enumerate}[label=(\roman*)]
    \item \textbf{Initialization.}
    An arbitrary initial value $\discrstep_j\left(0\right)$ is assigned at iteration $\historystep = 0$ around an arbitrary initial configuration $a_j$.
    There is no golden rule for selecting an appropriate starting value $\discrstep_j\left(0\right)$, and this choice represents yet another manifestation of the familiar exploitation-exploration trade-off inherent to optimization problems.
    The same initial value may be used for all variables.
    Intuitively, a large $\discrstep_j$ may prevent the algorithm from becoming trapped in local minima during the early stages of the optimization, although at the cost of slower convergence.
    \item \textbf{Update.}
    At iteration $\historystep = \historystep_k$, the configuration of $\nvars$ variables is
    \begin{equation*}
        \vec{x}\left(\historystep_k\right)
        =
        \left(
            x_1(\historystep_k),
            \, \ldots, \,
            x_j(\historystep_k),
            \, \ldots, \,
            x_{\scriptscriptstyle \nvars}(\historystep_k)
        \right),
    \end{equation*}
    with $\costfunc\left[\vec{x}\left(\historystep_k\right)\right] \equiv \costfunc_{\historystep_k}$.
    If the variable value remains constant after one iteration ($x_j(\historystep_{k + 1}) = x_j(\historystep_k)$), the discretization step of the $j$-th variable $x_j$ is updated according to
    \begin{equation}
        \label{eq:updating_rule}
        \begin{cases}
            \discrstep_j(\historystep_{k + 1})
            <
            \discrstep_j(\historystep_k)
            \quad \textit{if} \ \costfunc_{\historystep_{k + 1}} = \costfunc_{\historystep_k} \\
            \phantom{=} \\
            \discrstep_j(\historystep_{k + 1}) > \discrstep_j(\historystep_k)
            \quad \textit{if} \ \costfunc_{\historystep_{k + 1}} < \costfunc_{\historystep_k}
        \end{cases}.
    \end{equation}
    If instead $x_j(\historystep_{k + 1}) \neq x_j(\historystep_k)$, then $\discrstep_j$ is left unchanged.

    The rationale is the following.
    The parameter $\discrstep_j$ sets the local search scale of the algorithm for the variable $x_j$.
    When neither the variable nor the function value changes between two iterations, the algorithm decreases
    $\discrstep_j$, refining the search around the current value of $x_j$.
    Conversely, if  $x_j$ remains unchanged while the objective value decreases~---~due, for example, to an update of a different variable~---~the current value of $x_j$ is compatible with an improved configuration and $\discrstep_j$ is increased to allow exploration of a broader neighborhood.
    When $x_j(\historystep_{k + 1}) \neq x_j(\historystep_k)$, the step size is preserved since the update of the variable already provides new information at the appropriate search scale.
    \item \textbf{Refresh.}
    The value of $\discrstep_j$ for a given variable may be reset to an arbitrary value if it becomes smaller than a prescribed threshold.
    This \emph{refresh phase} prevents the algorithm from getting stuck in local minima due to an excessively small discretization step.
    Similarly, after an arbitrary number of iterations, all $\discrstep_j$ values may be restored to their initial settings, and the optimization may then be restarted from the best configuration obtained so far.
\end{enumerate}
Figure~\ref{fig:sub-region_tn} illustrates the evolution of the discretization step for three distinct variables over $140$ iterations of the algorithm.
The differing trends among the three variables reflect their individual dynamic adaptation during the optimization process.

\subsection{Local approximation model}\label{subsec:local_approx}

With the dynamically adapted discretization of
Section~\ref{subsec:dinamically-adapted}, Eq.~\eqref{eq:non_convex_problem} is solved by quantum or quantum-inspired methods by recasting $\costfunc$ as a many-qubit Hamiltonian and searching for its ground state~\cite{lucas2014_np-probs-as-ising}.

\begin{figure}[t]
    \centering
    \begin{overpic}[width=0.95 \columnwidth, unit=1mm]{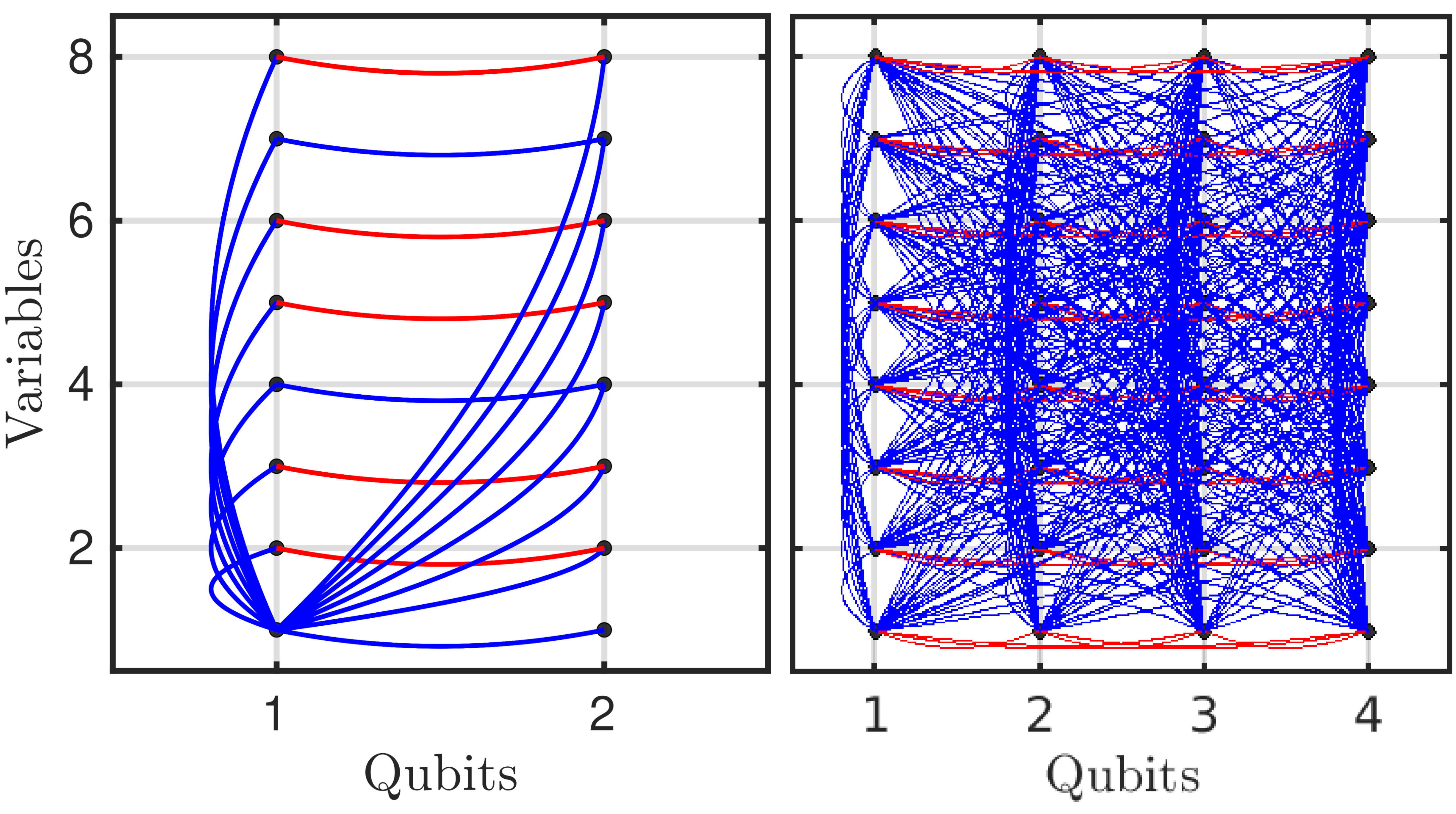}
    \end{overpic}
    \caption{
        \emph{Example of interaction graphs arising from the spin-glass mapping of a continuous-variable cost function.}
        Example of interaction graphs obtained from the local quadratic approximation around a given configuration $\vec{x}(\historystep_k)$ at iteration $\historystep_k$ as defined in Eq.~\eqref{eq:ising_model} during the optimization of the $8$-variable Gramacy-Lee~\cite{gramacy2012_glfunction} cost function, discretized using $\nqubits = 2$ (\textbf{left panel}) and $\nqubits = 4$ (\textbf{right panel}).
        Black dots represent qubits in the lattice, while blue and red lines denote, respectively, positive and negative long-range coupling strengths $J_{\ell\ell^\prime}$ among pairs of qubits.
    }
    \label{fig:interaction_schemes}
\end{figure}

Following and extending the approach introduced in Ref.~\cite{nazareth2015_dwave-quantum-annealing-IMRT}, we construct a spin-glass~\cite{edwards1975_ea-model, sherrington1975_sk-model, binder1986_spinglass, mezard1987_spinglass, wu1991_quantum-spinglass, nishimori2001_spinglass, charbonneau2023_spinglass} Hamiltonian around the current best configuration $\vec{x}(\historystep_k)$ at each iteration $\historystep_k$.
The Hamiltonian takes the form
\begin{equation}
    \label{eq:ising_model}
    \ham(\historystep_k)
    =
    \sum_{\ell=1}^{\nqubits \nvars} \sgbias_\ell(\historystep_k) \, \sgvar_\ell
    +
    \sum_{\ell=1}^{\nqubits \nvars} \sum_{\ell^\prime > \ell} \sgcoupling_{\ell\ell^\prime}(\historystep_k) \, \sgvar_\ell \sgvar_{\ell^\prime}
\end{equation}
where $\sgvar_{\ell} \in \left\{-1, 1\right\}$ denotes the spin (qubit) variable associated with site $\ell$. For  Eq.~\eqref{eq:ising_model} to serve as a surrogate of $\costfunc(\vec{x})$ around the configuration $\vec{x}(\historystep_k)$, the spin-glass couplings $\left(\left\{\sgbias_\ell(\historystep_k)\right\}, \left\{\sgcoupling_{\ell\ell^\prime}(\historystep_k)\right\}\right)$ must be appropriately determined.
The following procedure is repeated at each iteration $t_k$: 
\begin{enumerate}[label=(\roman*)]
    \item Starting from the configuration $\vec{x}(\historystep_k)$, we generate a set of neighboring configurations $\left\{\vec{x}_m(\historystep_k)\right\}_{m = 1}^{M}$. The cardinality $M$ is chosen to match the number of unknown coefficients in the surrogate Ising model, namely the local fields and the nonzero pairwise couplings.
    This number depends on the chosen number of qubits and equals
    \begin{equation*}
        M
        =
        \dfrac{\nvars\nqubits \, \left(\nvars \nqubits + 1\right)}{2}
    \end{equation*}
    for a fully-connected model.
    \item The function $F(\vec{x})$ is evaluated for each of these configurations, returning a set of $M$ values $\{F(\vec{x}_m(t_k))\}$.

    \item The Hamiltonian $\ham(\vec{x}_m(\historystep_k))$ is evaluated for each of these configurations using the representation in  Eq.~\eqref{eq:ising_model}.
    This yields a set of $M$ expressions $\{\ham(\vec{x}_m(t_k))\}$ where the unknown couplings $\left(\left\{\sgbias_\ell(\historystep_k)\right\}, \left\{\sgcoupling_{\ell\ell^\prime}(\historystep_k)\right\}\right)$ are free parameters.
    \item For each configuration $\vec{x}_m(\historystep_k)$, we enforce the matching condition between the corresponding value of the cost function $\costfunc(\vec{x}_m(\historystep_k))$ and expression $\{\ham(\vec{x}_m(t_k))\}$, 
    \begin{equation}
        \label{eq:equation_system}
        \costfunc(\vec{x}_m(\historystep_k))
        \overset{!}{=}
        \ham(\vec{x}_m(\historystep_k)),
    \end{equation}
    which yields a linear system in the unknown spin-glass couplings $\left(\left\{\sgbias_\ell(\historystep_k)\right\}, \left\{\sgcoupling_{\ell\ell^\prime}(\historystep_k)\right\}\right)$.
    Solving this system provides the set of coefficients that make the model in Eq.~\eqref{eq:ising_model} locally consistent (i.e. on the $M$ configurations sampled) with the true function over the sampling interval defined by our discretization scheme, i.e.,
    \begin{equation*}
        x^{(m)}_j
        \in
        \left[
            \refvalue_j - 2^{{\nqubits} - 1} \discrstep_j \, , \;
            \refvalue_j + \left(2^{\nqubits - 1} - 1\right) \discrstep_j
        \right],
    \end{equation*}
    where $x^{(m)}_j$ is the $j$-th component of the $m$-th configuration $\vec{x}_m(\historystep_k)$ generated in \emph{(i)} and discretized through Eq.~\eqref{eq:flexible_discretization}.
\end{enumerate} 
This mapping framework results in a \emph{local quadratic approximation}, see Eq.~\eqref{eq:ising_model}, of the original cost function $\costfunc$ around $\vec{x}(t_k)$. As previously noted, the resulting $\sgcoupling$ coupling matrix can exhibit all-to-all connectivity. In practice, however, sparser topologies may be required due to the specific connectivity of the different physical quantum processors. Figure~\ref{fig:interaction_schemes} illustrates two such interaction patterns produced by the local approximation procedure for the same configuration of the $8$-dimensional Gramacy-Lee~\cite{gramacy2012_glfunction} cost function (see Section~\ref{sec:results} for more details).
As shown, using a larger number of qubits to discretize the continuous variables leads to notably denser interaction graphs.
Therefore, the number of qubits per variable must be carefully chosen depending on each problem instance. 
This procedure is repeated at every iteration $\historystep_k$.
Candidate solutions to the original optimization problem in Eq.~\eqref{eq:non_convex_problem} are then obtained by sampling low-energy configurations of the Hamiltonian $\ham(t_k)$. This is the stage where the quantum(-inspired) solver is employed.
At the end of each iteration $\historystep_k$, the new candidate solutions generated by the solver are evaluated using the objective function $\costfunc$, and the one with the lowest value $\costfunc(\vec{x}_{best}(\historystep_k))$ is selected.

More precisely, after each iteration $\historystep_k$, two sets of configurations are available:
\begin{enumerate}
    \item ${\mathcal{N}(\historystep_k)} = \left\{\vec{x}_m(\historystep_k)\right\}_{m = 1}^{M}$, which is the set of neighboring configurations used to construct the surrogate model;
    \item ${\mathcal{Q}(\historystep_k)}$, which is the set of configurations produced by the quantum (or quantum-inspired) solver.
\end{enumerate}
To control the computational cost of our optimization scheme, we monitor the total number of function evaluations $\funcevaluation \in \mathbb{N}$, and impose the stopping condition
$\funcevaluation > \maxfuncevaluation$, where $\maxfuncevaluation \in \mathbb{N}$ is the maximum allowed number of evaluations.
This stopping criterion is particularly suitable when comparing different solvers, as it is solver-agnostic.
In contrast, a stopping condition based on a fixed number of iterations may lead to unfair comparisons, since the meaning of iteration can vary significantly across solvers.

\subsection{Tensor-network optimization}\label{subsec:tensor_network}

Tensor networks (TNs) are variational representations of quantum many-body states in which the exponentially large tensor coefficient of a wavefunction is decomposed into a network of smaller tensors connected according to a chosen geometry~\cite{orus2014_tn, montangero2018_tn}.
Their effectiveness relies on the fact that many physically relevant low-energy states occupy only a limited corner of the full Hilbert space and can therefore be described accurately with a comparatively small number of parameters~\cite{vidal2003_slightly-entangled, hastings2007_area-law, eisert2010_area-law}.
In practice, the expressive power of a TN ansatz is controlled by the bond dimension $\bonddimension \in \mathbb{N_{+}}$, which quantifies the amount of entanglement retained in the representation and at the same time determines the numerical cost of the simulation. 

In one spatial dimension, the paradigmatic TN ansatz is the matrix product state \mbox{(MPS)~\cite{schuch2008_mps, schollwock2011_dmrg-tn-connection-and-mps-review, silvi2019_anthology, cirac2021_mps-peps-review}}, which also provides the modern variational interpretation of the density-matrix renormalization group \mbox{(DMRG)~\cite{white1992_dmrg, verstraete2004_dmrg-tn-connection, mcculloch2007_dmrg-tn-connection, schollwock2011_dmrg-tn-connection-and-mps-review}}.
Other tensor-network geometries, such as projected entangled-pair states \mbox{(PEPS)~\cite{schuch2006_peps, verstraete2006_peps, cirac2021_mps-peps-review}}, tree tensor networks \mbox{(TTNs)~\cite{shi2006_ttn, tagliacozzo2009_ttn-2d, murg2010_ttn-higher-dimension, silvi2010_ttn, silvi2019_anthology, montangero2018_tn}}, and multiscale entanglement renormalization Ans\"atze \mbox{(MERA)~\cite{vidal2007_mera, corboz2009_mera, evenbly2009_mera}}, extend the same general philosophy to more structured correlation patterns and higher-dimensional settings.

Historically, TN methods were developed in close connection with strongly correlated quantum many-body physics~\cite{vidal2004_mps-time_evolution, white2004_dmrg, rico2014_tn-4-lgt, cirac2021_mps-peps-review, verstraete2023_dmrg, magnifico2025_tn-4-lgt}, where they became standard tools for the study of ground states, low-lying excitations, and static properties of lattice Hamiltonians.
Over time, their range of applications has broadened considerably, and TNs are now also widely used as classical emulators and benchmarking tools for quantum algorithms whenever the entanglement generated during the computation remains sufficiently controlled~\cite{zhou2020_tn-vs-qc, collura2024_tn}.
From this perspective, tensor networks offer a flexible environment to test new mappings and encodings before targeting quantum hardware directly.

In the present work, TNs are employed precisely in this spirit, namely as a classical numerical tool to optimize the effective spin-glass surrogate Hamiltonian emerging from the mapping introduced in the previous Sections.
Among the possible tensor-network geometries, we choose the TTN ansatz because its hierarchical structure is naturally suited to capture long-range correlations and to represent efficiently interaction patterns beyond a simple one-dimensional chain~\cite{silvi2019_anthology, montangero2018_tn}.
This feature is particularly appealing in our setting, where the surrogate Hamiltonians resulting from Eq.~\eqref{eq:equation_system} are characterized by glassy all-to-all interactions.
Moreover, TTNs have recently shown promise in addressing hard classical optimization problems encoded as classical spin-glass systems~\cite{cavinato2021_radiotherapy-tn, tesoro2026}.

The optimization is therefore formulated as a variational ground-state search and carried out through a DMRG-like sweeping procedure implemented within the \texttt{qtealeaves} library~\cite{quantumtea, qtealeaves}.
Once the variational low-energy state of the surrogate Hamiltonian has been prepared, candidate solutions are extracted by sampling bit-strings from its low-energy sector via tensor-network sampling techniques~\cite{ballarin2025_opes}.

\subsection{Benchmark solvers}\label{subsec:benchmark_solvers}

To assess the performance of the proposed optimization framework, and in particular of the TTN-based solver, we compare it with three reference strategies:
\begin{enumerate}[label=(\roman*)]
    \item \textbf{Best Neighbor (BN)}.
    To isolate the contribution of the surrogate-Hamiltonian optimization step, we introduce a simple baseline referred to as the Best Neighbor (BN) strategy.
    At each iteration, BN selects the configuration with the lowest value of the original objective function among the neighboring points used to construct the local surrogate model, without invoking any additional optimization solver.
    This benchmark therefore quantifies the improvement provided by solving the surrogate problem, rather than relying solely on the information already contained in the local neighborhood.
    \item \textbf{Simulated Annealing (SA)}.
    As an alternative solver for the same surrogate Hamiltonian, we employ the \texttt{SimulatedAnnealingSampler} provided in the \texttt{dwave-ocean-sdk}~\cite{kirkpatrick1983_sa, dwave-ocean_sa}.
    At each iteration of the outer optimization loop, SA is applied to the discretized surrogate model to generate a set of low-energy candidate configurations.
    In the present work, the number of reads is set to $10{,}000$, allowing us to sample multiple candidate solutions at each optimization step.
    This provides a direct comparison with the TTN-based solver on the same encoded problem.
    \item \textbf{Particle Swarm (PS)}.
    To further benchmark our method against a well-established classical metaheuristic, we consider Particle Swarm (PS) optimization~\cite{eberhart1995_particle-swarm}.
    PS is a population-based, derivative-free optimization algorithm inspired by the collective behavior of biological swarms.
    It evolves a set of candidate solutions, referred to as \emph{particles}, which explore the search space by combining individual and collective search directions in order to approach high-quality solutions.
    Its use as a benchmark is motivated by two main features: it does not require gradient information, making it suitable for non-differentiable landscapes, and it is naturally amenable to parallel implementations.
    In this work, its performance is mainly controlled by the swarm size, denoted by $SZ$, which fixes the total number of particles.
    We use the implementation provided in the Global Optimization Toolbox of \texttt{MATLAB R2022a}~\cite{matlab2026_particle-swarm}.
\end{enumerate}
Both BN and SA operate on the same dynamically discretized local surrogate representation used by the TTN-based solver.
Their comparison with TTN therefore probes the relative effectiveness of different optimization strategies on the same encoded problem.
By contrast, PS is applied directly to the original continuous objective function, providing an external benchmark against a standard classical optimizer acting in the native continuous search space.

\section{Results}
\label{sec:results}

This Section reports the numerical results obtained with the proposed dynamically adapted discretization and local spin-glass surrogate Hamiltonian framework introduced in Section~\ref{sec:M&M}.

We begin with the unconstrained $n$-dimensional Gramacy--Lee benchmark, comparing the TTN-based solver with BN and SA on the same discretized representation.
We then compare our method with PS optimization, consider a constrained variant of the benchmark, and finally apply the framework to a non-polynomial inverse-planning problem in fixed-gantry intensity-modulated radiotherapy (IMRT).

\subsection{Unconstrained optimization problem}\label{subsec:unconstrained_optimization_problem}

\begin{figure}[t]
    \centering
    \begin{overpic}[width=0.7 \columnwidth, unit=1mm]{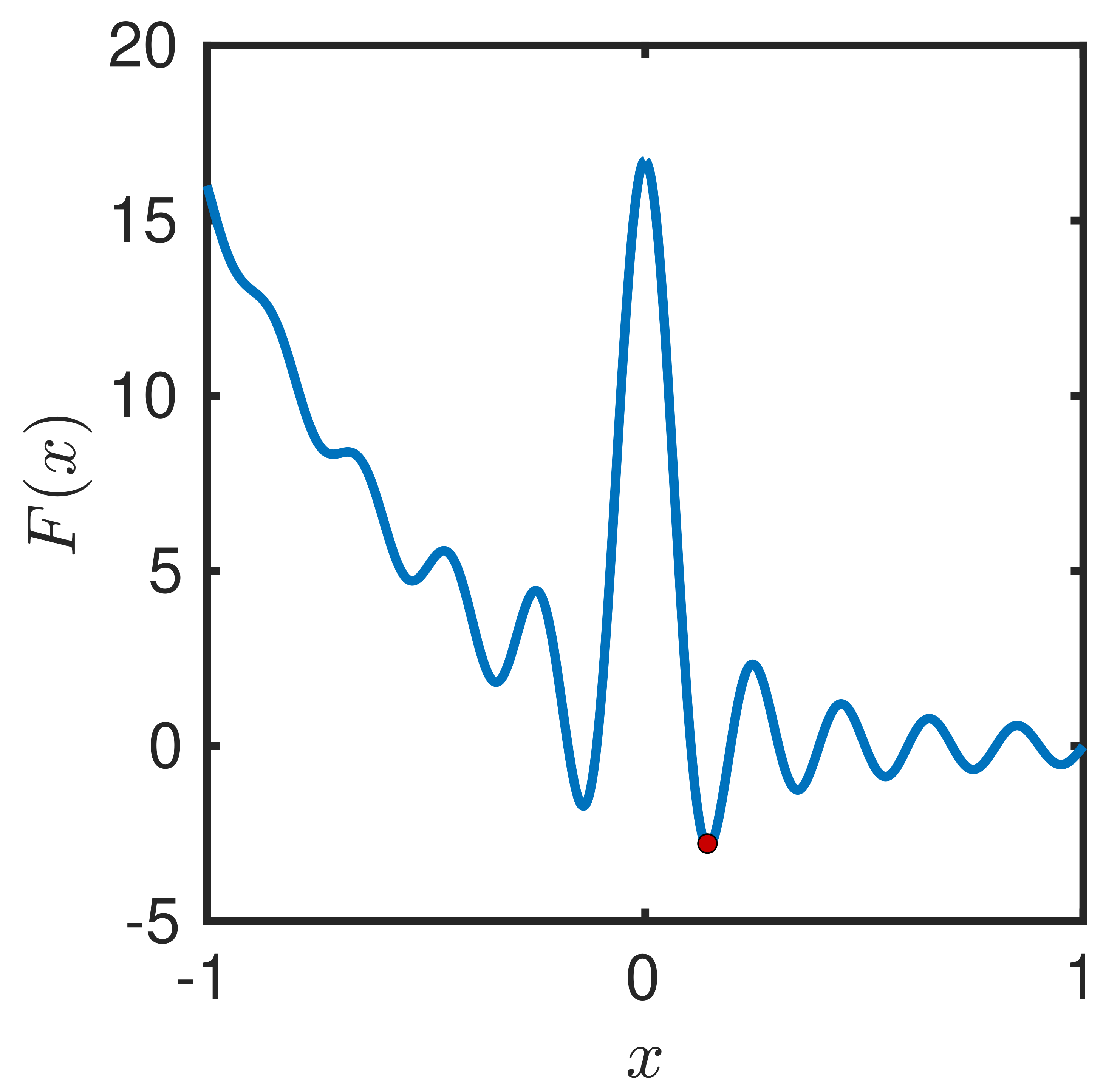}
    \end{overpic}
    \caption{
        \emph{One-dimensional Gramacy-Lee function.}
        The Figure plots an instance of the one-dimensional Gramacy-Lee function~\cite{gramacy2012_glfunction} as defined in Eq.~\eqref{eq:GLN} for $w = 10$ in the domain $x \in \left[-1, 1\right]$.
        The global minimum $x_\star \simeq 0.14$ is marked by a red point.
    }
    \label{fig:GL}
\end{figure}

To validate the proposed approach, we first consider the unconstrained minimization of the \emph{$n$-dimensional Gramacy-Lee ($n$GL)} function~\cite{gramacy2012_glfunction}
\begin{equation}
    \label{eq:GLN}
    \costfunc(\vec{x})
    =
    \sum_{j=1}^n
    \dfrac{\sin\left(\pi \, w_j \, x_j\right)}{2x_j}
    +
    \left(x_j - 1\right)^4 ,
\end{equation}
with parameters $w_j > 0$ for all $j = 1, \ldots, n$.
The $n$GL function features a highly non-convex landscape with multiple local minima, and it is widely used as a benchmark for testing optimization algorithms~\cite{surjanovic2013_opt-test-functions}.
Figure~\ref{fig:GL} illustrates the one-dimensional case $n = 1$ of Eq.~\eqref{eq:GLN} for $w_1 \equiv w = 10$~\cite{surjanovic2013_opt-test-functions}.
Within the interval $x \in [-1,1]$, the global minimum is located approximately at $x_\star \simeq 0.14$, and the corresponding $n$-dimensional minimizer is therefore $\vec{x}_{\star} \simeq \left(0.14, 0.14, \ldots, 0.14\right)$.
Because of the $1 \mathbin{/} x_j$ prefactor, the analytic expression in Eq.~\eqref{eq:GLN} has a removable singularity at $x_j = 0$, which gives rise to the sharp feature visible near the origin.
At the same time, its oscillatory and non-polynomial form does not admit a natural direct low-order binary encoding of the type assumed in standard QUBO/PUBO formulations~\cite{kochenberger2014_qubo-survey, lucas2014_np-probs-as-ising, glover2022_quantum-bridge-analytics, mandal2020_pubo1}.
These properties make the $n$GL family a suitable and demanding benchmark for assessing the proposed continuous-variable mapping strategy.

The optimization problem in Eq.~\eqref{eq:GLN} is solved for increasing numbers of variables $n \equiv \nvars \in \left\{8, 16, 32, 64\right\}$.
For all cases, we set the number of qubits per variable to $\nqubits = 2$. Moreover, for each value of $n$, we conduct $\nsamples = 10$  independent optimization runs. Each run $r \in \{1,..., 10\}$ in the sample starts from a unique and randomly initialized configuration of the problem variables $\vec{x}_r(\historystep_0)$, and proceeds independently of the others. To ensure a fair comparison between solvers, the same set of random starting points is used for all solvers at a given value of $\nvars$.
Following the stopping criterion discussed at the end of Section~\ref{subsec:local_approx}, we impose a limit on the maximum number of function evaluations,
\begin{equation*}
    {\maxfuncevaluation}
    =
    150 \times M ,
\end{equation*}
where the second factor corresponds to the number of configurations required to construct the local surrogate model under a fully connected interaction scheme.
Consequently, at most $150$ iterations are performed when the sampler fails to identify candidate solutions throughout the optimization run, in which case only neighboring solutions are evaluated.
With this procedure, the resulting limits on the number of function evaluations are $\maxfuncevaluation \in \left\{2.04 \times 10^4, 8 \times 10^4, 3 \times 10^5, 1.2 \times 10^6\right\}$ for $n \equiv \nvars \in \left\{8, 16, 32, 64\right\}$, respectively.

Unless otherwise stated, the bond dimension of the TTN in this work simulations is set to $\bonddimension = 5$, and the shaded regions shown in the following Figures represent $95\%$ confidence intervals for the sample mean, computed using the standard Student-$t$ prescription appropriate for a finite number of runs~\cite{NIST_ehandbook}.

\begin{figure}[t]
    \centering
    \begin{overpic}[width=0.95 \columnwidth, unit=1mm]{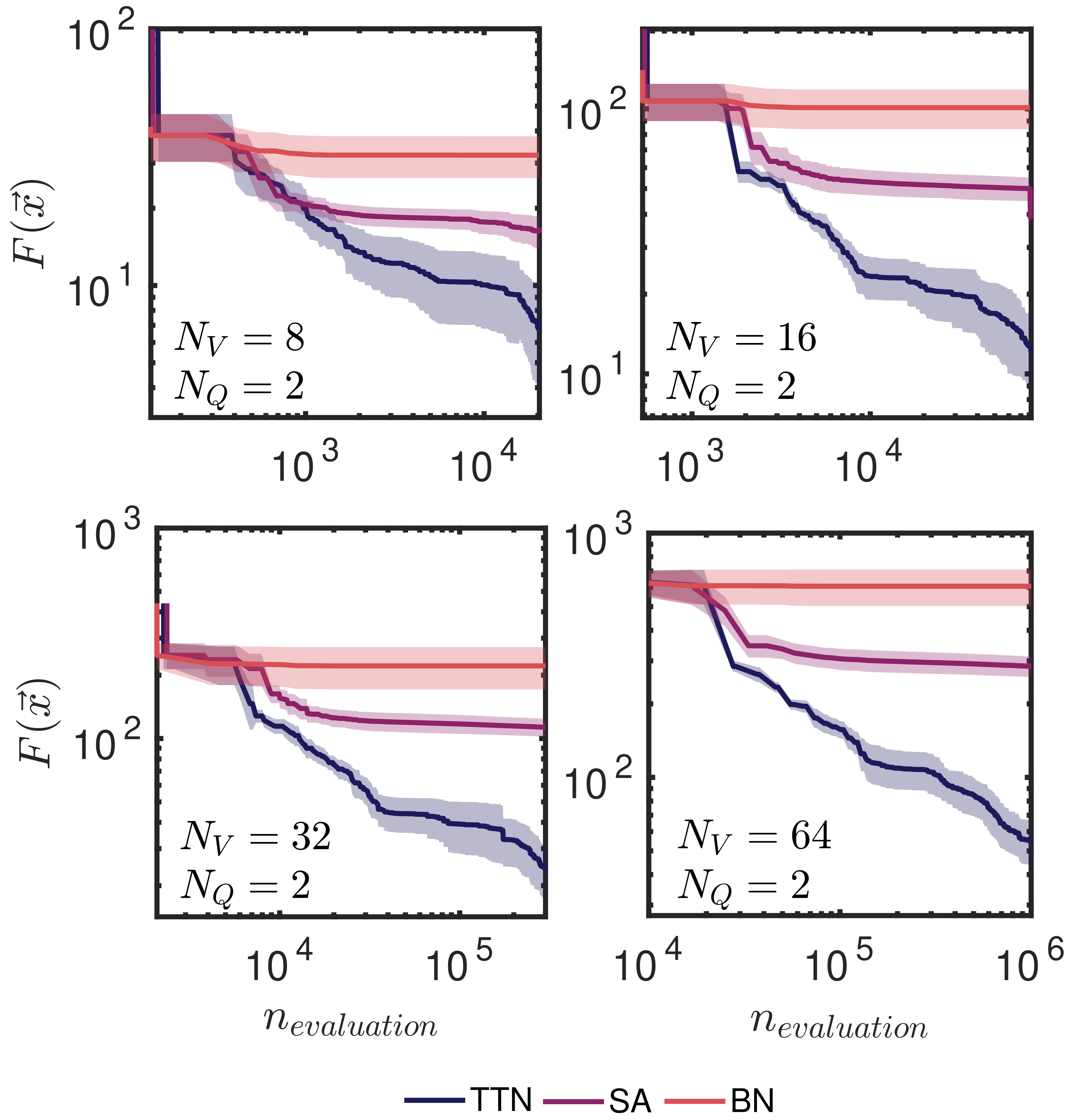}
    \end{overpic}
    \caption{
        \emph{Optimization of the \mbox{$n$-dimensional} \mbox{Gramacy-Lee} function with $\nqubits = 2$.}
        Evolution of the objective value $F(\vec{x})$ defined in Eq.~\eqref{eq:GLN} as a function of the number of function evaluations $\funcevaluation$, shown on logarithmic scales for $\nvars \in \{8,16,32,64\}$.
        Each panel corresponds to a different problem size, with $\nqubits = 2$ qubits per variable in all cases.
        Blue, magenta, and red curves refer respectively to the Tree Tensor Network (TTN), Simulated Annealing (SA), and Best Neighbor (BN) solvers.
        Solid lines denote the sample mean over $\nsamples = 10$ independent optimization runs, while the shaded regions indicate the corresponding $95\%$ confidence intervals.
    }
    \label{fig:2qubits_comparison}
\end{figure}

Figure~\ref{fig:2qubits_comparison} shows the evolution of the objective function during the optimization for different numbers of variables and for the various solvers considered.
The performance of the TTN and SA solvers is compared against that of BN.
Among the three approaches, TTN consistently achieves the best performance, both in terms of convergence speed and final objective value.
Moreover, the advantage of TTN becomes increasingly pronounced as the number of variables $\nvars$ grows, as indicated by the widening gap between the TTN and SA curves.
The steeper slope observed for TTN relative to SA further suggests that additional improvement could be expected if the optimization were allowed to proceed for a larger number of iterations.
In contrast, the BN algorithm becomes trapped in a local minimum during the early stages of the optimization, leading to markedly inferior performance.

This behavior highlights the effectiveness of both TTN and SA in exploring the solution space beyond local (neighboring) configurations and in identifying higher-quality solutions.
Both TTN and SA outperform BN from this perspective, with TTN providing the best overall performance in the present benchmark.

\begin{figure}[t]
    \centering
    \begin{overpic}[width=0.95 \columnwidth, unit=1mm]{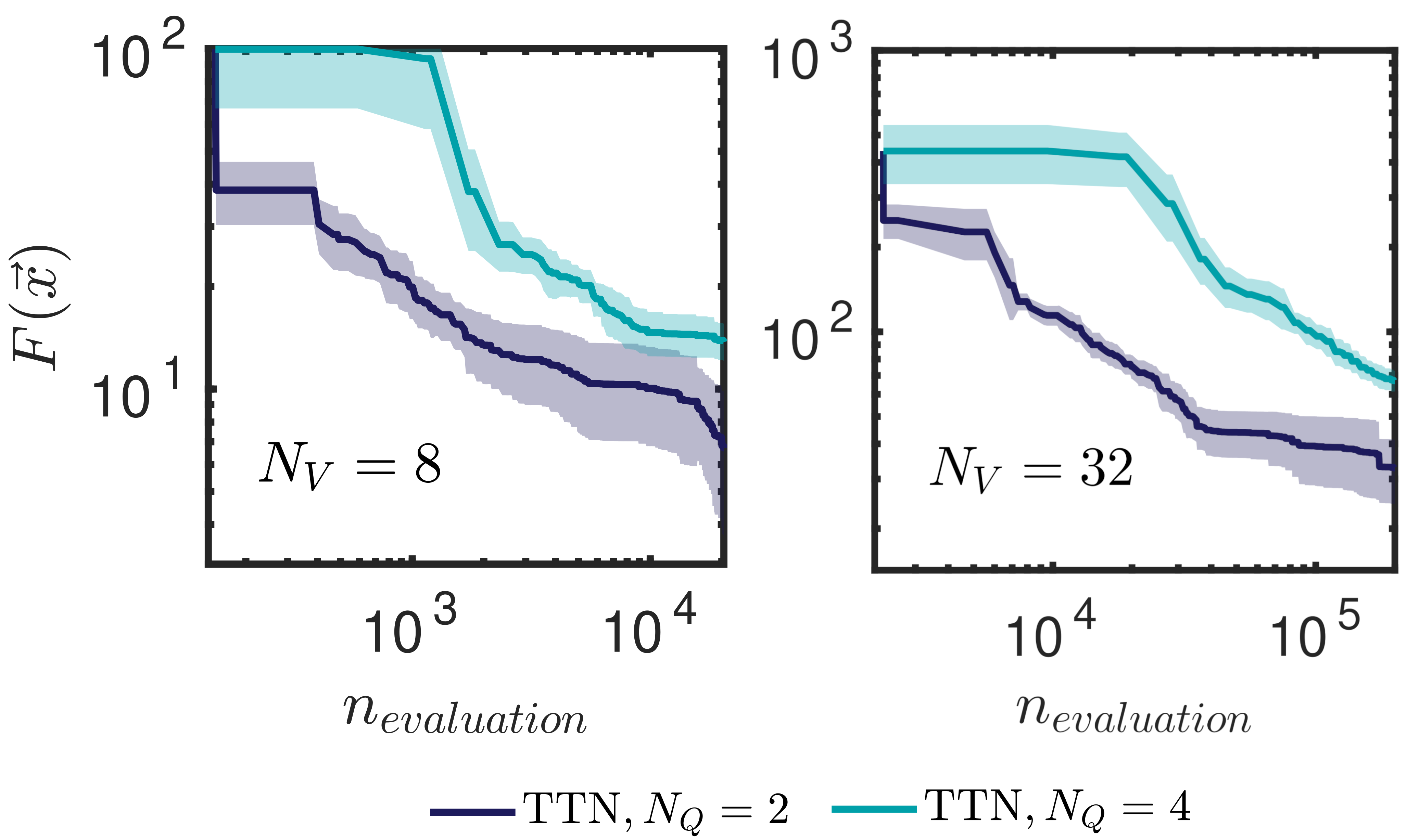}
    \end{overpic}
    \caption{
        \emph{TTN-based optimization of the \mbox{$n$-dimensional} \mbox{Gramacy-Lee} function with different number of qubits $\nqubits$.}
        Comparison of the TTN-based results obtained for the $n$GL function with $\nvars = 8$ and $\nvars = 32$, using $\nqubits = 2$ and $\nqubits = 4$ qubits per variable in the discretization.
        The plots show the evolution of the objective value $F(\vec{x})$ defined in Eq.~\eqref{eq:GLN} as a function of the number of function evaluations $\funcevaluation$, on logarithmic scales.
        Solid lines denote the sample mean over $\nsamples = 10$ independent optimization runs, while the shaded regions indicate the corresponding $95\%$ confidence intervals.
    }
    \label{fig:2vs4qubits_comparison}
\end{figure}

Figure~\ref{fig:2vs4qubits_comparison} compares the results obtained using $\nqubits = 2$ and $\nqubits = 4$ qubits per variable for the same $n$GL optimization with $\nvars = 8$ and $\nvars = 32$ variables. As shown in the Figure, increasing the number of qubits per variable to $\nqubits = 4$ does not lead to improved solutions when compared at an equal number of function evaluations. 
This behavior is expected.
Indeed, a larger value of $\nqubits$ requires a greater number of neighboring configurations to construct the local surrogate model, an effect that is particularly pronounced under the fully connected interaction scheme employed here.
As a consequence, the neighborhood $\mathcal{N}(\historystep_k)$ evaluated at each iteration is larger for $\nqubits = 4$ than for $\nqubits = 2$, and thus, for a fixed total number of evaluations $\funcevaluation$, a larger fraction of the evaluations is spent on neighboring configurations when $\nqubits$ is increased.

Taken together, these results indicate that the proposed approach is intrinsically well suited to operate with small values of $\nqubits$, while efficiently scaling to larger problem sizes $n \equiv \nvars$.

\subsection{Comparison with the Particle Swarm algorithm}\label{subsec:comparison_with_PS}

We repeat the benchmark in Section~\ref{subsec:unconstrained_optimization_problem}, this time comparing our method with the \mbox{\emph{Particle Swarm (PS)}~\cite{eberhart1995_particle-swarm, matlab2026_particle-swarm}} optimization algorithm.
We assess the performance of the two approaches as a function of the problem dimensionality, considering $n \equiv \nvars \in \{8,16,32,64\}$, with $\nqubits = 2$ and $\nsamples = 10$.

To ensure a fair and controlled comparison between PS and our method, we adopt the following protocol.
The swarm size (SZ) is chosen to match the number of configurations used to construct the local surrogate model in our approach.
In addition, the “particles” are initialized at the same positions as the configurations employed to build the initial local quadratic approximation.
This procedure guarantees that both algorithms start from identical initial conditions.

Finally, the comparison is carried out using the same number of function evaluations, $\funcevaluation$, following the same rationale adopted in the previous benchmarks.

\begin{figure}[t]
    \centering
    \begin{overpic}[width=0.95 \columnwidth, unit=1mm]{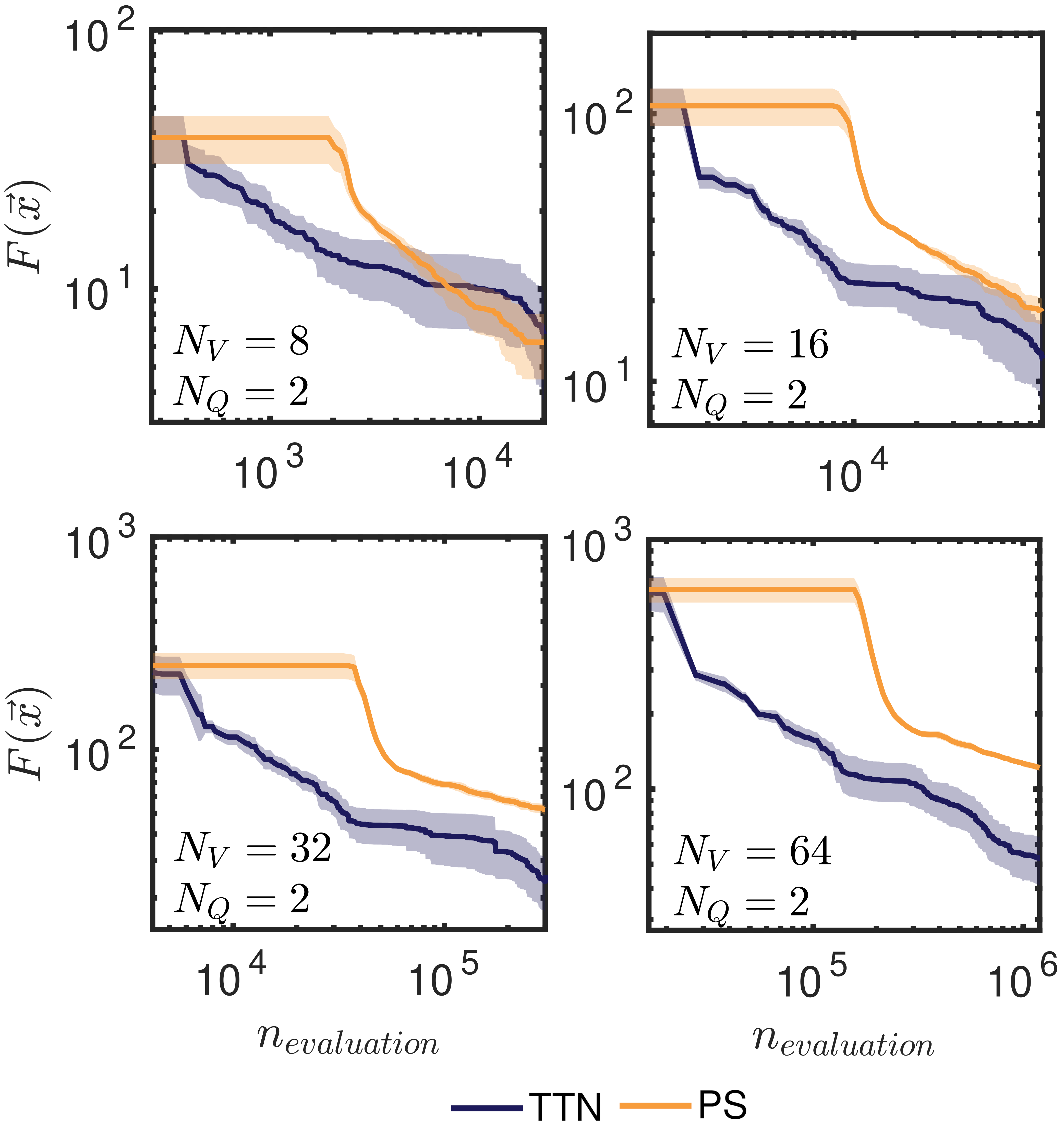}
    \end{overpic}
    \caption{
        \emph{Comparison between the TTN-based solver and Particle Swarm optimization for the \mbox{$n$-dimensional} Gramacy--Lee function.}
        Evolution of the objective value $F(\vec{x})$ defined in Eq.~\eqref{eq:GLN} as a function of the number of function evaluations $\funcevaluation$, shown on logarithmic scales for $\nvars \in \{8,16,32,64\}$.
        Each panel corresponds to a different problem size, with $\nqubits = 2$ qubits per variable used in the TTN-based discretization.
        Blue and orange curves refer, respectively, to the Tree Tensor Network (TTN)-based solver and to Particle Swarm optimization (PS).
        Solid lines denote the sample mean over $\nsamples = 10$ independent optimization runs, while the shaded regions indicate the corresponding $95\%$ confidence intervals.
        Unlike the TTN-based solver, which acts on the discretized surrogate Hamiltonian, PS is applied directly to the original continuous objective function.
    }
    \label{fig:ttn_ps_comparison}
\end{figure}

The results are presented in Figure~\ref{fig:ttn_ps_comparison}.
A clear qualitative difference between the two algorithms emerges, particularly during the early stages of the optimization.
In general, PS exhibits a pronounced initial stagnation phase, followed by a rapid convergence, whereas our method displays a more consistent and faster convergence from the outset. For the smallest problem size, $n \equiv \nvars = 8$, this difference does not significantly affect the quality of the final solution.
As the dimensionality increases, however, the performance gap progressively widens in favor of the TTN-based approach.
We emphasize that this behavior may depend on the specific problem under consideration and should not be interpreted as universally applicable.
Nevertheless, these results highlight the effectiveness of our proposed optimization strategy when compared with a well-established classical algorithm such as PS.
We further note that all results reported here are obtained using $\nqubits = 2$.

\subsection{Constrained optimization}\label{subsec:constrained_optimization}

\begin{figure}[t]
    \centering
    \begin{overpic}[width=0.65\columnwidth, unit=1mm]{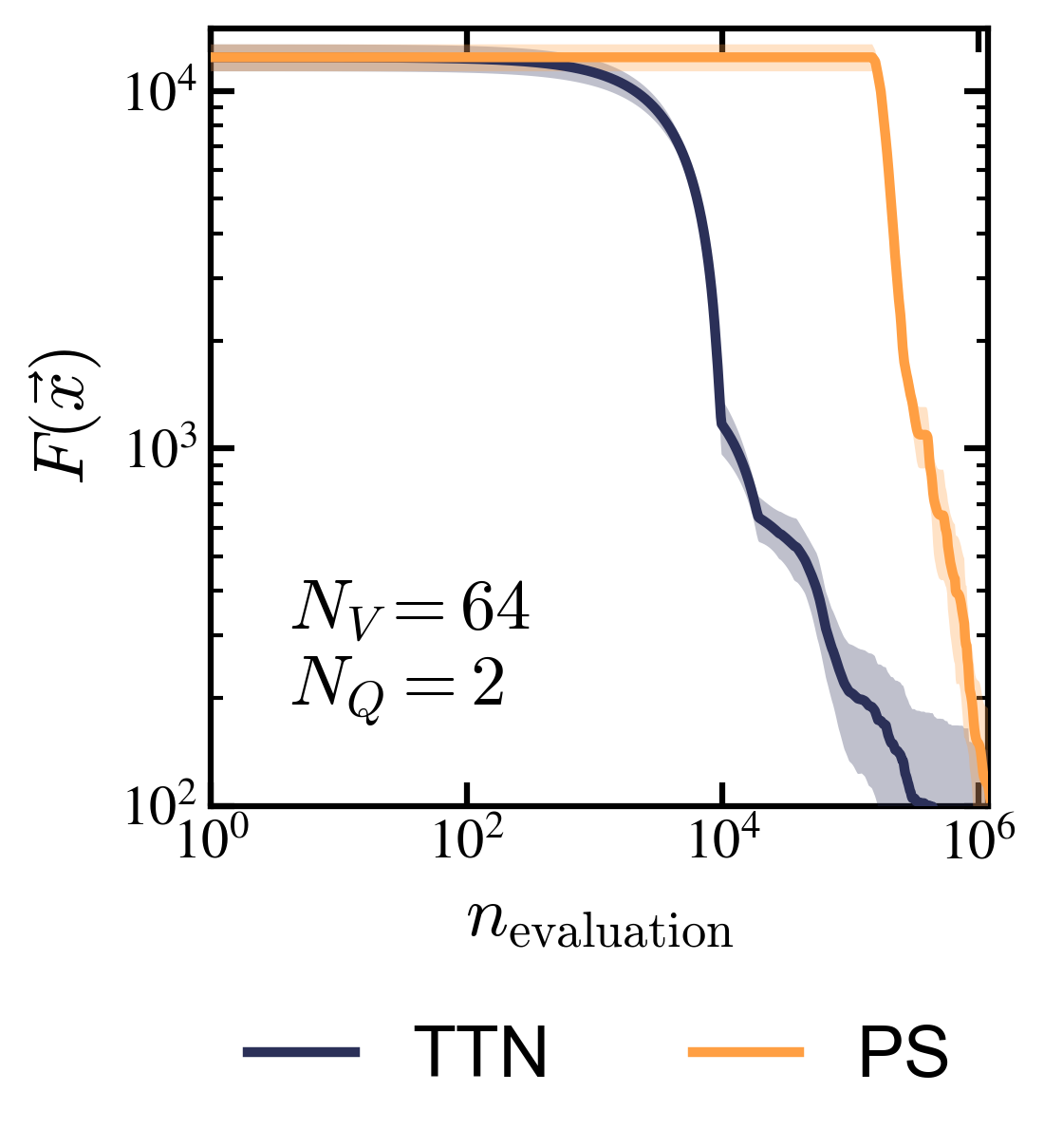}
    \end{overpic}
    \caption{\emph{Comparison between the TTN-based solver and Particle Swarm optimization for the constrained \mbox{$n$-dimensional} Gramacy--Lee function.}
        Evolution of the constrained objective function defined in Eq.~\eqref{eq:constrained_GLN} with $\mathcal{K}=200$ as a function of the number of function evaluations $\funcevaluation$, shown on logarithmic scales for the case $\nvars = 64$.
        The TTN-based results are obtained using $\nqubits = 2$ qubits per variable in the discretization strategy.
        Blue and orange curves refer, respectively, to the TTN-based solver and to Particle Swarm optimization (PS).
        Solid lines denote the sample mean over $\nsamples = 10$ independent optimization runs, while the shaded regions indicate the corresponding $95\%$ confidence intervals.
    }
    \label{fig:GLN_constrained}
\end{figure}

To further assess the robustness of our approach, we next consider a constrained optimization problem obtained by modifying Eq.~\eqref{eq:GLN} as
\begin{equation}
    \label{eq:constrained_GLN}
    \costfunc(\vec{x})
    =
    \sum_{j = 1}^n
    \dfrac{\sin\left(\pi \, w_j \, x_j\right)}{2x_j}
    +
    \left(x_j - 1\right)^4
    +
    \mathcal{K} \sum_{j = 1}^n \dfrac{x_j}{\left\lvert{x_j}\right\rvert} .
\end{equation}
The additional term introduces a sign-dependent penalty: it increases the objective function for positive values of the variables and decreases it for negative ones, thereby systematically favoring negative configurations over positive ones.
As a consequence, the global minimum of the objective function is shifted from its original location to approximately $\vec{x}_\star \simeq \left(-0.14, -0.14, \ldots, -0.14\right)$, which constitutes the new optimum of the constrained problem.
The parameter $\mathcal{K} \in \mathbb{R}_{+}$ controls the strength of this penalty term and thus the extent of the constraint.
Figure~\ref{fig:GLN_constrained} reports a comparison between the TTN-based approach and the PS algorithm on the constrained problem for $n \equiv \nvars = 64$.
All the other simulation hyper-parameters used for this constrained benchmark are the same as in the previous analyses.
The results confirm the effectiveness of our method also in the presence of constraints, showing that it achieves solutions of comparable quality to those obtained using PS.

\subsection{Application to radiotherapy inverse-planning}\label{subsec:imrt_opt_problem_nonconvex}

As a third application to further validate our methodology, we consider a realistic optimization problem arising in radiotherapy treatment planning~\cite{taylor2004_imrt}.
In particular, we focus on \emph{Intensity-modulated Radiotherapy (IMRT)}~\cite{bortfeld2006_imrt}, a standard-of-care technique characterized by two defining features:
\begin{enumerate}
    \item the spatial modulation of radiation beam intensity (\emph{fluence}), which is not spatially uniform;
    \item the determination of the final treatment plan through an optimization procedure commonly referred to as \emph{inverse planning}.
\end{enumerate}
The central task of inverse planning is to determine an optimal fluence map that maximizes dose delivery to the tumor while simultaneously minimizing irradiation of the surrounding healthy tissue.
This requirement gives rise to a large-scale and highly structured optimization problem that must be solved within clinically acceptable time constraints. The objective of IMRT is to generate a patient-specific dose distribution that delivers the prescribed dose to the target volume while sparing the organs at risk (OARs).
To this end, each radiation beam is discretized into a grid of narrow \emph{pencil beams}, or \emph{beamlets}.
Only beamlets intersecting the target are included in the optimization, and the intensity of the $j$-th beamlet is independently controlled by a non-negative weight $x_j \in \mathbb{R}_{+}$.

The patient anatomy is discretized into three-dimensional finite-volume elements, or \emph{voxels}.
The total dose deposited in the $i$-th voxel, denoted by $\dosedistribution_i$, is obtained as the sum of contributions from all beamlets, leading to the global dose distribution $\dosedistribution(\vec{x})$.
This relationship is expressed as
\begin{equation}
    \label{eq:dose_per_voxel}
    \dosedistribution_i
    =
    \sum_{j=1}^{\nbeamlets} \influencemel_{ij} \, x_j \;,
\end{equation}
where $\nbeamlets$ is the total number of beamlets and $\influencemel_{ij}$ are the elements of the influence matrix $\influencematrix$, which encodes the unmodulated contribution of the $j$-th beamlet to the $i$-th voxel.

In this work, we formulate the inverse planning problem through the composite cost function
\begin{equation}
    \label{eq:inverse_planning_problem_1}
    \costfunc(\vec{x})
    =
    \costfunc_{\scriptscriptstyle \mathrm{P}}(\vec{x})
    +
    \costfunc_{\scriptscriptstyle \mathrm{NP}}(\vec{x}) \; ,
\end{equation}
where the subscripts $\mathrm{P}$ and $\mathrm{NP}$ denote \emph{polynomial} and \emph{non-polynomial} contributions, respectively.
The polynomial term is defined as
\begin{equation}
    \label{eq:inverse_planning_problem_2}
    \costfunc_{\scriptscriptstyle \mathrm{P}}(\vec{x})
    =
    \sum_{r=1}^{\mathrm{R}}
    \sum_{i=1}^{\nvoxel_r}
    \gamma_i
    \left[\dosedistribution_i(\vec{x}) - \dosedistribution_i^{(r)} \right]^2 \; ,
\end{equation}
where the index $r$ runs over all anatomical structures (target tumor and OARs), $\gamma_i \in \mathbb{R}$ is a voxel-dependent weight used to prioritize specific regions, and $\nvoxel_r \in \mathbb{N}$ denotes the total number of voxels belonging to structure $r$. Here, $R$ is the total number of structures involved, with $R = 3$ in our case.

The non-polynomial contribution incorporates maximum/minimum dose constraints and is given by
\begin{equation}
    \label{eq:inverse_planning_problem_3}
    \begin{split}
        \costfunc_{\scriptscriptstyle \mathrm{NP}}(\vec{x})
        =
        &{\underbrace{\sum_{r=1}^{\mathrm{R}} \alpha_r \left[\max \left(0, \, \dosedistribution^{(r)}(\vec{x}) - \dosedistribution_{\scriptscriptstyle\mathrm{max}}^{(r)}\right)\right]^2}_{\mathrm{A}}} \; + \\
        & + {\underbrace{\sum_{r=1}^{\mathrm{R}} \beta_r \left[\min \left(0, \, \dosedistribution^{(r)}(\vec{x}) - \dosedistribution_{\scriptscriptstyle \mathrm{min}}^{(r)}\right)\right]^2}_{\mathrm{B}}} \; ,
    \end{split}
\end{equation}
where the two terms enforce complementary penalties: term $\mathrm{A}$ penalizes voxels in structure $r$ that exceed the maximum tolerated dose $\dosedistribution_{\scriptscriptstyle \mathrm{max}}^{(r)}$, weighted by $\alpha_r \in \mathbb{R}$, while term $\mathrm{B}$ penalizes voxels that fail to reach the prescribed minimum dose $\dosedistribution_{\scriptscriptstyle \mathrm{min}}^{(r)}$, with importance factor $\beta_r \in \mathbb{R}$. In addition, since beamlet intensities must be non-negative, an explicit penalty ($\mathcal{K}=500$) enforcing this constraint is included in the total cost function, following the strategy validated in Section~\ref{subsec:constrained_optimization}.

\begin{figure}[t]
    \centering
    \begin{overpic}[width=1.0 \columnwidth, unit=1mm]{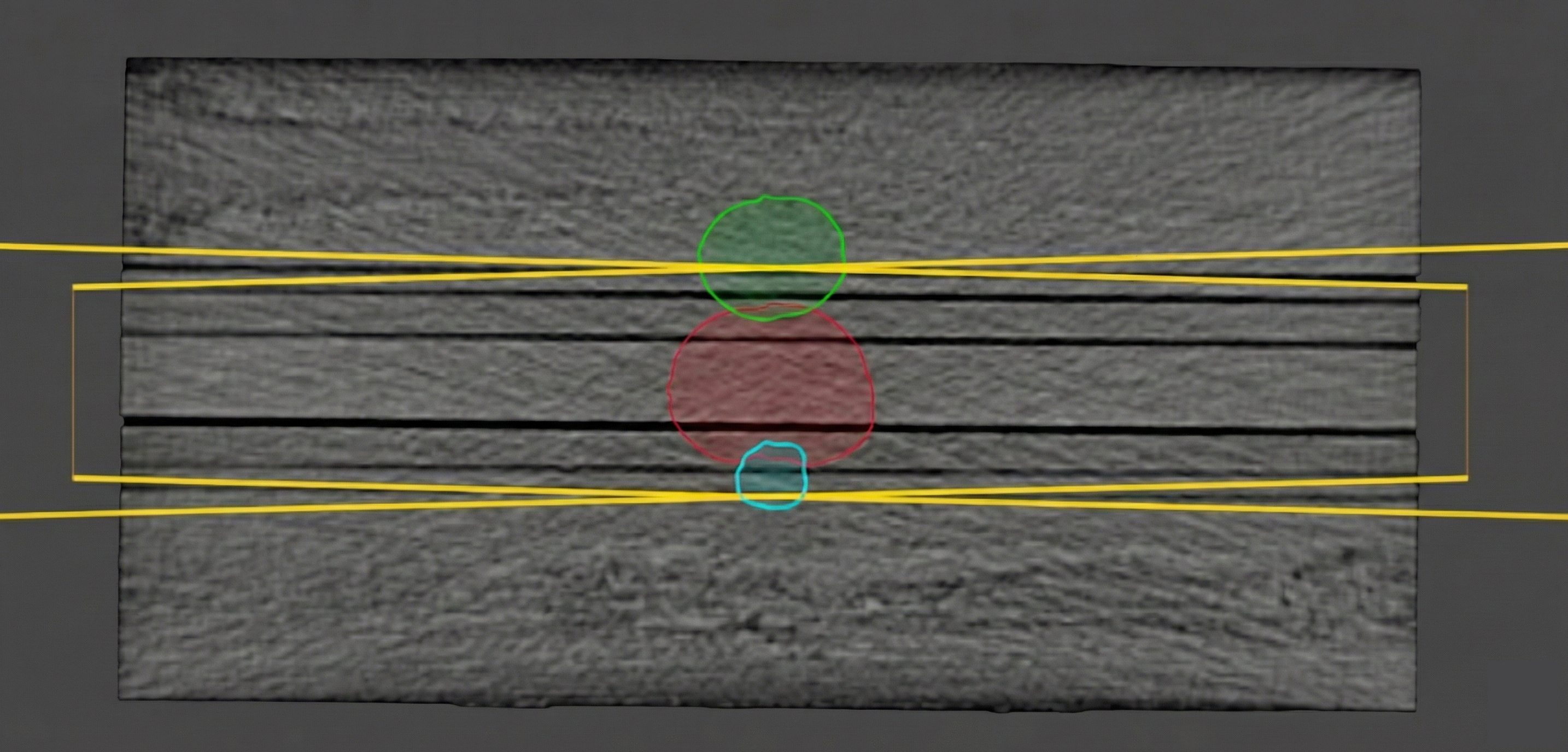}
    \end{overpic}
    \caption{
        \emph{TG$-119$ phantom, optimization structures and beams arrangement.}
        Representation of an axial slice of the 3D Computed Tomography (CT) image of the TG-$119$ phantom. The yellow lines represent the two opposing treatment beams (at 90$^\circ$ and 270$^\circ$). The three structures used in the optimization are outlined in green (Bladder), red (Planning Target Volume - PTV~\cite{vanherk2004_radio}) and light blue (Rectum).
        }
    \label{fig:tg_119_phantom}
\end{figure}

Consistent with a previous study of some of the authors on the quadratic polynomial formulation in Ref.~\cite{cavinato2021_radiotherapy-tn}, we consider the prostate case of the TG-$119$ phantom.
The radiation beam geometry consists of two opposing radiation fields at $90^\circ$ and $270^\circ$, as illustrated in Figure~\ref{fig:tg_119_phantom}. Beams were subdivided into a total of $\nbeamlets = 64$ beamlets.
Each beamlet weight was discretized using two qubits, resulting in a total of $\nvars \!\times\! \nqubits = 64 \!\times\! 2 = 128$ qubits.
Three anatomical structures were considered. They are reported in Table~\ref{tab:non_convex_constraints}, together with the optimization objectives and associated priorities.
\begin{table}[t]
    \centering
    \begin{tabular}{p{3cm} p{2.5cm} p{1.5cm}}
        \midrule
        \midrule
        \textbf{Structure} & \textbf{Objective} & \textbf{Priority} \\
        \midrule
        \midrule
        \multirow{3}{*}{PTV} 
            & $\dosedistribution_i = 1.0$ 
            & $\gamma = 1.8$ \\
            & $\dosedistribution_{\scriptscriptstyle \mathrm{\max}} = 1.07$ 
            & $\alpha = 1.5$ \\
            & $\dosedistribution_{\scriptscriptstyle \mathrm{\min}} = 0.98$ 
            & $\beta = 1.5$ \\
        \midrule
        \multirow{2}{*}{Rectum} 
            & $\dosedistribution_i = 0.0$ 
            & $\gamma = 1.2$ \\
            & $\dosedistribution_{\scriptscriptstyle \mathrm{\max}} = 0.8$ 
            & $\alpha = 1.2$ \\
        \midrule
        \multirow{2}{*}{Bladder} 
            & $\dosedistribution_i = 0.0$ 
            & $\gamma = 1.2$ \\
            & $\dosedistribution_{\scriptscriptstyle \mathrm{\max}} = 0.8$ 
            & $\alpha = 1.2$ \\
        \midrule
        \midrule
    \end{tabular}
    \caption{
        \emph{Inverse-planning optimization problem objectives and priorities.}
        List of objectives and priorities used in this work for the optimization of the non-polynomial inverse planning problem defined in Eq~\eqref{eq:inverse_planning_problem_3} on the TG-$119$ phantom prostate case. 
    }
    \label{tab:non_convex_constraints}
\end{table}
We emphasize that these parameters are not derived from clinical treatment protocols; rather, they are chosen to demonstrate the ability of the algorithm to handle non-polynomial objective functions.
This choice is further motivated by the reduced number of beamlets compared to full-scale clinical scenarios.
\begin{figure*}[t]
    \centering
    \begin{overpic}[width=1.0\textwidth, unit=1mm]{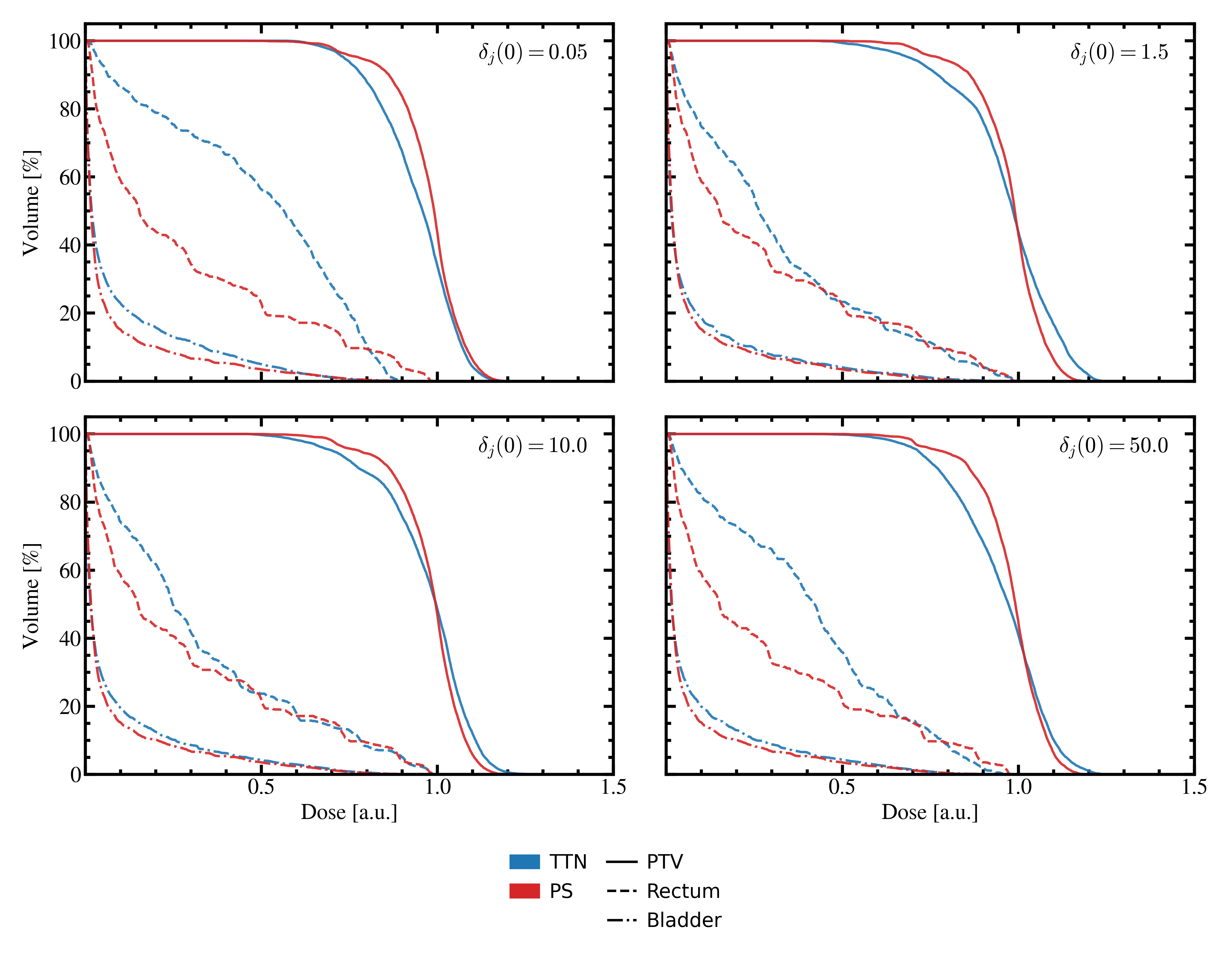}
    \end{overpic}
    \caption{
        \emph{Comparison of the DVHs of three structures obtained with TTN-based and PS optimization.}
        Comparison between the DVHs of the three structures (PTV, Rectum and Bladder) obtained by using the TTN-based and the PS optimization algorithms. For each run, the
swarm size (SZ) is chosen to match the number of configurations used to construct the local surrogate model in
our approach. In addition, the “particles” are initialized
at the same positions as the configurations employed to
build the initial local quadratic approximation.}
    \label{fig:non_convex_dvh_a}
\end{figure*}

Figure~\ref{fig:non_convex_dvh_a} compares the solutions obtained using the TTN-based approach and the PS algorithm, for different values of the initial discretization step $\delta_j(0)$. As previously explained, this value influences the set of initial beamlet values used to create the approximation model. Furthermore, it also defines the set of initial configurations used by the PS algorithm.  Therefore, both methods are initialized identically and subject to the same limits on function evaluations.

The optimization results are presented in terms of \emph{Dose–Volume Histograms (DVHs)}, a standard tool for radiotherapy plan evaluation.
DVHs report the fraction of a given anatomical structure (i.e., Planning Target Volume (PTV), Rectum or Bladder in this case) receiving at least a specified dose level. The two algorithms run for a maximum of $150$ iterations, with an upper limit of $\funcevaluation = 1.2 \!\times\! 10^6$ function evaluations.
The optimization is initialized with all beamlet weights set to zero.
Globally, the two algorithms exhibit comparable solutions in terms of DVH.
 The best solutions are achieved with $\discrstep_j(0) = 10$, yielding cost function values of $1033$ and $755$ for TTN and PS, respectively. The most significant discrepancies between the two algorithms regarding DVH are observed for the rectum when $\delta_j(0) = 0.05$ and $\delta_j(0) = 50$. In both instances, the solution provided by TTN exhibits a lower $\dosedistribution_{\scriptscriptstyle \mathrm{max}}$, although it underperforms at lower dose levels. PS proved to be extremely stable and reproducible, with values falling within the range $[754, 758]$ across all runs. TTN also demonstrated stability, albeit with values distributed over a broader range $[1033, 1387]$. The higher cost values observed in TTN compared to PS may be attributed to the discretization of the problem, specifically concerning the approximation method for the model construction. However, this does not undermine the validity of the approach. Indeed, the provided solutions were very close to those obtained using a classical benchmark on the original continuous objective function, and were obtained by using the newly proposed discretization strategy with two qubits per variable.

\begin{figure}[t] 
    \centering
    \begin{overpic}[width=1.0\columnwidth, unit=1mm]{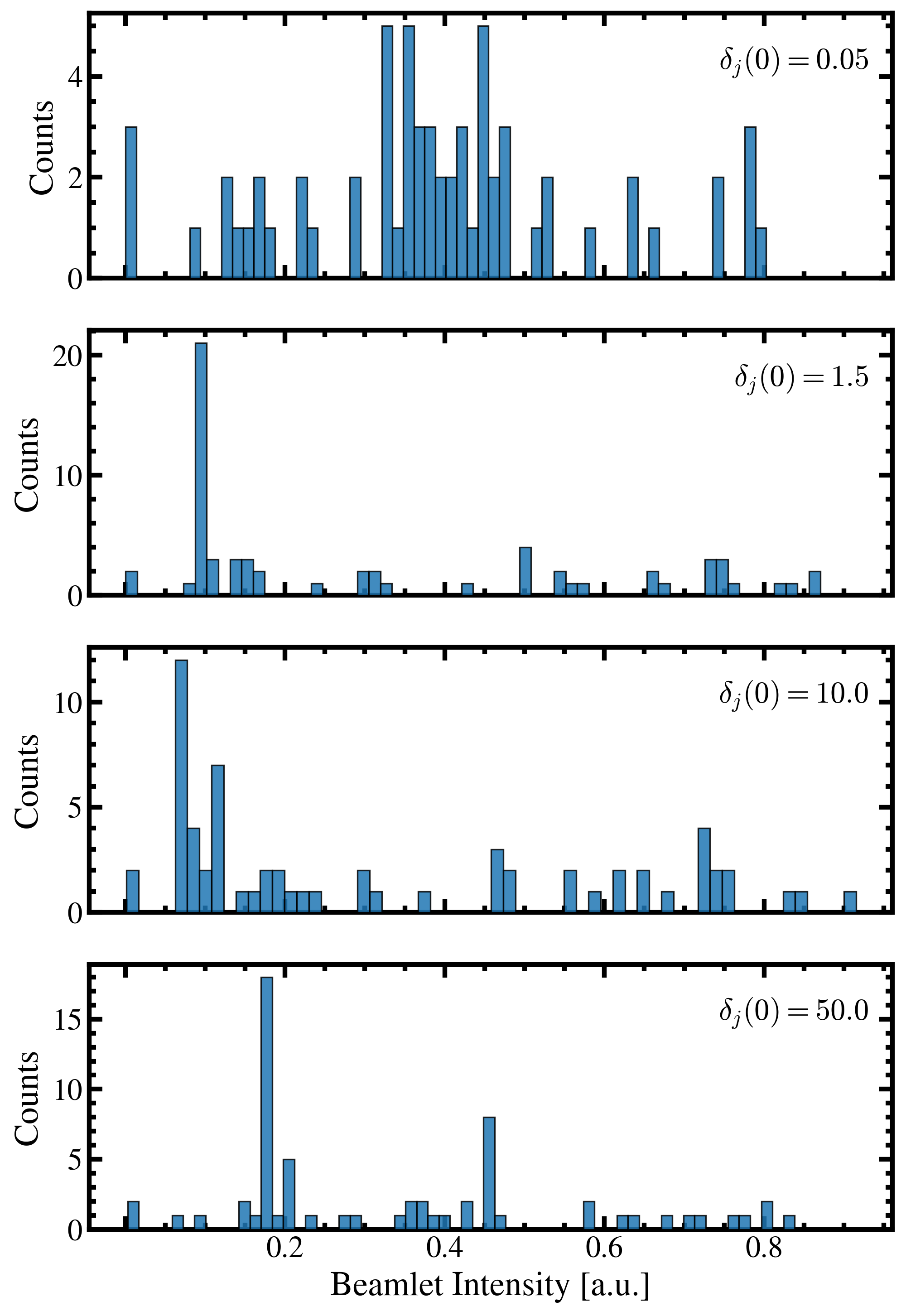}
    \end{overpic}
    \caption{\emph{Beamlet-weights values obtained with the TTN-based approach.}
        Comparison between the beamlet-weights values explored by the TTN-based approach during the optimization for the four values of $\delta_j(0)$ considered.
    }
    \label{fig:beamlet_distribution}
\end{figure}

This aspect is further underscored by the data presented in Figure~\ref{fig:beamlet_distribution}, which depicts the full set of $x_j$ values explored by the TTN-based approach during optimization for each $\delta_j(0)$ value. As can be observed, the resulting distributions exhibit a continuum of values. This contrasts with the discrete distribution that would result from a standard binary-to-decimal conversion, which would be limited to only $4$ distinct values  distributed over a fixed range, when employing $\nqubits = 2$ qubits per variable. Consequently, this provides further empirical evidence—beyond the purely conceptual rationale—of the flexibility inherent in our approach for optimizing continuous problems on discrete resources such as quantum computers.
Furthermore, the TTN-based approach exhibited a faster cool-down rate during the initial optimization stages compared to PS, similarly to what has already been shown in Sections~\ref{subsec:comparison_with_PS}~and~\ref{subsec:constrained_optimization} on a different problem.
This highlights the potential advantage of non-local exploration offered by our approach from the very beginning.

\section{Conclusions}
\label{sec:conclusions}

In this work, we have introduced an approach for the optimization of general functions defined over continuous variables, targeting implementation on quantum hardware. 
The framework combines two main ingredients. 
The first is a dynamically adapted discretization scheme, which maps each continuous variable to a small register of qubits while updating the local resolution during the optimization. 
This allows the algorithm to sample different discretized neighborhoods during the search, rather than being restricted to a fixed grid. 
The second is a local surrogate construction, which maps the objective function, independently of its analytic form, onto an effective spin-glass Hamiltonian around the current configuration. 
The resulting Hamiltonian is then optimized using a TTN-based DMRG-like variational ground-state search.

We tested this strategy on three classes of problems: the unconstrained $n$-dimensional Gramacy--Lee benchmark, a constrained variant of the same benchmark, and a proof-of-concept fixed-gantry IMRT inverse-planning problem. 
In the benchmark cases, the TTN-based implementation consistently improved over the Best Neighbor baseline and showed favorable performance with respect to Simulated Annealing on the same discretized representation. 
When compared with Particle Swarm optimization, which acts directly on the original continuous problem, the proposed method achieved competitive results and, in several cases, faster initial convergence.

The fixed-gantry IMRT application tested the method on a structured optimization problem with both polynomial dose-matching terms and non-polynomial penalty contributions. 
The TTN-based approach produced dose-volume histograms close to those obtained with Particle Swarm optimization while using only $\nqubits = 2$ qubits per beamlet weight. 
However, the current implementation did not outperform Particle Swarm in terms of the final cost-function value.  
These results should therefore be interpreted as a proof of principle for the flexibility of the encoding and surrogate-construction strategy, rather than as a demonstration of clinical-scale performance. Future work should focus on scalability, including larger IMRT instances, optimized tensor-network strategies, alternative surrogate solvers, and reduced-connectivity Hamiltonian constructions. 
These developments would support applications to more realistic treatment-planning problems and to advanced modalities such as Volumetric-Modulated Radiotherapy (VMAT)~\cite{otto2008_vmat} and Helical Tomotherapy~\cite{mackie2006_tomotherapy}.

Overall, the proposed framework provides a compact mapping strategy for continuous-variable optimization on quantum and quantum-inspired platforms. 
Its main advantage is that arbitrary objective functions, including non-polynomial, non-differentiable, or numerically defined costs, can be incorporated through a local spin-glass surrogate without requiring a global polynomial binary reformulation of the original problem.


\begin{acknowledgments}

This work is partially supported by the European Union via \mbox{ICSC - Italian} Research Center on HPC, Big Data and Quantum Computing (NextGenerationEU Project No. CN00000013), project PASQuanS2 (Quantum Technology Flagship), project QEC4QEA  (HORIZON-EUROHPC-JU-2023-QEC-05 Project No. 101194322); the Italian Ministry of University and Research (MUR) via: Quantum Frontiers (the Departments of Excellence 2023-2027); the World Class Research \mbox{Infrastructure - Quantum} Computing and Simulation Center (QCSC) of Padova University; Istituto Nazionale di Fisica Nucleare (INFN): iniziativa specifica \mbox{IS-QUANTUM}; the program PON “Ricerca e Innovazione” 2014-2020 (project 19-G-12542); the QuanTEdu-France program (State grant part of France 2030, project No. ANR-22-CMAS-0001)
We acknowledge computational resources from Cloud Veneto, CINECA, and the University of Padova Strategic Research Infrastructure Grant 2017: “CAPRI: Calcolo ad Alte Prestazioni per la Ricerca e l’Innovazione”.

\end{acknowledgments}


\clearpage
\bibliographystyle{apsrev4-2}  
\bibliography{references}  

\end{document}